\documentclass[]{JHEP3} 

\usepackage{epsfig,multicol,bbm}

\newcommand\fverb{\setbox\fverbbox=\hbox\bgroup\verb}
\newcommand\fverbdo{\egroup\medskip\noindent%
			\fbox{\unhbox\fverbbox}\ }
\newcommand\fverbit{\egroup\item[\fbox{\unhbox\fverbbox}]}
\newbox\fverbbox

\title{Mesons at large $N_c$ from lattice QCD}

\author{Gunnar S.\ Bali\\
Institut f\"ur Theoretische Physik, 
Universit\"at Regensburg, 93040 Regensburg, Germany\\
	E-mail: \email{gunnar.bali@physik.uni-regensburg.de}}
\author{Francis Bursa\\
Rudolf Peierls Centre for Theoretical Physics,
University of Oxford,\\
1 Keble Road, Oxford OX1 3NP, U.K.\\
E-mail: \email{f.bursa1@physics.ox.ac.uk}}

\preprint{}

\abstract{We calculate the pion and $\rho$ meson masses in quenched
$\mathop{\rm SU}(N)$ gauge theories for $N=2, 3, 4$ and $6$. Extrapolating
these results to the chiral and large-$N$ limits, we find
$m_{\rho}=(1.670\pm 0.024 )\sqrt{\sigma}$
for the $\rho$ meson mass
at a fixed lattice spacing $a\approx 0.2093\,\sigma^{-1/2}\approx0.093$~fm,
where we use the (arbitrary) value ($444$~MeV$)^2$ for the string tension.
We estimate a continuum limit large-$N$ value, 
$m_{\rho}=(1.77\pm 0.05)\sqrt{\sigma}$.
We find $1/N^2$ corrections to be small and we compare
our results to predictions from AdS/QCD.
}


\begin{document} 


\section{Introduction}
Numerical simulations of QCD in lattice regularization
(Lattice QCD) are the standard way
to derive low energy aspects of the phenomenology of strongly
interacting elementary particles, directly from
the QCD Lagrangian. Adding conformal invariance via
supersymmetry or sending the number
of colours $N$ from three to infinity are popular ways
of turning some aspects of non-perturbative QCD analytically tractable.
Such extensions are in particular also necessary to
connect supergravity predictions to properties of induced
four dimensional quantum field theories on the boundary
of anti-de Sitter space~\cite{Maldacena} (AdS/CFT correspondence or
AdS/QCD~\cite{Gursoy1,Gursoy2,Erdmenger}).

Pure $\mathop{\rm SU}(N)$ gauge theories have
been studied extensively
on the lattice and, among other quantities,
the glueball spectra, string tensions~\cite{Lucini:2004my},
$\overline{MS}$ $\Lambda$-parameters~\cite{Lucini:2008vi}
and de-confinement transition temperatures~\cite{Lucini:2003zr}
are well-established. These pure Yang-Mills
results demonstrate
that $1/N^2$ corrections are already below the
10~\% level at $N=3$.
Little is known as yet for full QCD with finite $N>3$, including $n_F$
flavours of sea quarks,
in which case corrections to the meson spectrum
are expected to be only suppressed by factors $n_F/N$.

Here, as a step in this direction, we study
the meson spectrum in the quenched approximation for
$N=2, 3, 4$ and $6$. While these finite-$N$ results alone will not allow
us to quantify the difference between $N=3$ and the
large-$N$ limit of full QCD, the quenched and the full theory
will share the same large-$N$ limit where sea quark loops are
suppressed for any finite number of quark flavours~\cite{Hooft}.

Obtaining large-$N$ results that are close to existing full
$N=3$ QCD results
(or indeed to real world experiment) will help us to understand
why the quenched approximation (and even the
naive quark model) works remarkably well in many
cases~\cite{Aoki:1999yr}. The large-$N$ expansion is also a powerful
tool to reduce the number of low energy constants in
chiral effective field theories~\cite{Scherer:2002tk} and we wish to understand
in what cases \emph{three} can be regarded as a large number and under
what circumstances not. Moreover, QCD in the planar
limit is an interesting quantum field theory in itself.
Last but not least, only in this limit the AdS/QCD correspondence
can work.

First exploratory
simulations of the pion masses in
$\mathop{\rm SU}(17)$ and $\mathop{\rm SU}(19)$ were
performed by Kiskis, Narayanan and Neuberger~\cite{Kiskis:2002gr}
some time ago, employing overlap fermions.
Preliminary results for the $\pi$ and $\rho$ meson masses
using the Wilson gauge and fermion actions for $N=2,3,4$ and 6 were
presented by us at Lattice 07~\cite{Bali:2007kt}. The present
article concludes this study.
Recently, Del Debbio \emph{et al.}~\cite{DelDebbio:2007wk},
obtained the $\pi$ and $\rho$ meson masses as well,
using the same action at the same $N$-values,
at a somewhat coarser lattice spacing than ours.
This coincidence gives us some control of the continuum limit.

This article is organised as follows: in section~\ref{sec:methods}
we introduce the methods employed in the simulation.
In section~\ref{sec:results} we describe our analysis methods
and present the results. We discuss finite volume corrections,
the $N$-dependence of the additive quark mass renormalization
and the $\rho$ meson mass as a function of the $\pi$ mass
at large $N$. In section~\ref{sec:compare} we relate our results to those
of Ref.~\cite{DelDebbio:2007wk} and perform the continuum limit extrapolation,
before we conclude the article with a discussion in section~\ref{sec:discuss}.

\section{Simulation parameters and lattice methods}
\label{sec:methods}
We use the software package
\emph{Chroma}~\cite{Chroma},
to carry out our simulations which we have adapted to work
for a general number of colours $N$.

We use the Wilson plaquette gauge action,
\begin{equation}
S_g=\beta\sum_{\Box}\left(1-\frac{1}{N}\mathrm{Re}\,\mathrm{Tr}\,U_{\Box}\right)\,,
\end{equation}
where $U_{\Box}$ is the product of
$\mathop{\rm SU}(N)$ link matrices $U_{x,\mu}$, connecting the
site $x$ with the site $x+a\hat{\mu}$, around an
elementary plaquette $\Box$. $a$ denotes the lattice spacing and
$\hat{\mu}$ a unit vector in $\mu$-direction.
We use Wilson fermions with hopping parameter $\kappa$,
\begin{equation}
S_f=\sum_x \psi^{\dagger}_x\psi_x - \kappa \sum_{x,\mu}\left[
\psi^{\dagger}_x(1-\gamma_\mu)U_{x,\mu}\psi_{x+a\hat{\mu}} +
\psi^{\dagger}_x(1+\gamma_\mu)U^\dagger_{x-a\hat{\mu},\mu}\psi_{x-a\hat{\mu}}\right]\,.
\end{equation}
$\kappa$ is related to the lattice quark mass $m_q$ by
\begin{equation}
am_q=\frac{1}{2}\left(\frac{1}{\kappa}-\frac{1}{\kappa_c}\right)
\quad\mbox{and}\quad\kappa_c^{-1}=8+{\mathcal O}(\beta^{-1})\,.
\label{bare mass}
\end{equation}

To set the scale, we use the string tension calculations by Lucini
\emph{et al.}~\cite{Lucini:2005vg}. We choose the coupling
$\beta=2N/g^2=2N^2/\lambda$
such that the infinite-volume string tension in lattice units
$a\sqrt\sigma$ is the same for each $N$.
We use the value $a\sqrt\sigma\approx 0.2093$: for $\mathop{\rm SU}(3)$
this corresponds
to $\beta=6.0175$.
$\lambda$ denotes the 't~Hooft coupling in the lattice scheme.
At our lattice spacing it converges towards the
$N\rightarrow\infty$ value $\lambda\approx 2.78$.
Adopting the value $\sigma=1$~GeV/fm~$\approx (444$~MeV$)^2$
for the string tension, the lattice spacing is $a\approx 0.093$~fm
in each case. The values of $a\sqrt\sigma$~\cite{Lucini:2005vg} used in the
fits are very accurate~\cite{Lucini:2004my},
so the mismatch 
of our lattice spacings in units of the string
tension between different $N$ is below the 1~\% level.

We have chosen our quark masses (or equivalently our values of $\kappa$)
in order to match the pion masses $m_{\pi}/\sqrt{\sigma}$
between the different $N$-values as closely as possible.
These $\kappa$-values were estimated by carrying out short
exploratory simulations of $m_{\pi}$ at different mass points
and interpolating.

\TABULAR{|c|c|c|c|l|c|}
{\hline
$N$ & $\beta$ & $\lambda=2N^2/\beta$ &Volume & $\kappa$ & $n_{\rm conf}$ \\ \hline
2 & 2.4645 & 3.246&$12^3\times 32$ & 0.1510 & 100 \\
& & &$16^3\times 32$ & 0.1457, 0.1480, 0.1500, 0.1510 & 100 \\
& & &$24^3\times 32$ & 0.1510 & 100 \\ \hline
3 & 6.0175 &2.991& $12^3\times 32$ & 0.1547 & 50 \\
& & &$16^3\times 32$ & 0.1500, 0.1520, 0.1537, 0.1547 & 50 \\ \hline
4 & 11.028&2.902 & $12^3\times 32$ & 0.15625 & 50 \\
& & &$16^3\times 32$ & 0.1520, 0.1540, 0.1554, 0.15625 & 50 \\ \hline
6 & 25.452 &2.829& $12^3\times 32$ & 0.15715 & 50 \\
& && $16^3\times 32$ & 0.1525, 0.1550, 0.1563, 0.15715 & 44 \\ \hline
}
{Simulation parameters. $n_{\rm conf}$ denotes the number
of independent gauge configurations.
\label{parameters}}

Our simulation parameters are given in table~\ref{parameters}.
We carry out most of our calculations on lattices with volume
$16^3\times32$ in lattice units, corresponding to a
spatial extent of $\approx 1.5$ fm in physical units. We carry out additional
simulations on $12^3\times32$ lattices at our lightest quark masses,
where finite size effects (FSE) are expected to be largest.
In $\mathop{\rm SU}(2)$ we augment this by a $24^3\times32$ volume, since
FSE are also expected to become larger with smaller $N$.
Indeed, FSE are expected to be \emph{zero} at infinite $N$, as long
as the lattice is kept larger than a critical length $l_c$~\cite{Kiskis:2003rd}.
Hence an extrapolation at a fixed finite volume to infinite
$N$ will yield the correct $N=\infty$ limit, in spite of any residual
FSE we may encounter at finite $N$.

We calculate correlators using three sources:
`point', `narrow' and `wide', and the corresponding three sinks.
We employ Jacobi-Gauss smearing~\cite{Gauss}, 
\begin{equation}
\psi^\prime_x=\psi_x+
\gamma\sum_i\left(U_{x,i}\psi_{x+a\hat{\imath}}+
U^{\dagger}_{x-a\hat{\imath},i}\psi_{x-a\hat{\imath}}\right),
\label{eq:gauss}
\end{equation}
with smearing parameter $\gamma=4$, with 10 iterations for
the `narrow' sources and sinks and 60 for the `wide' ones.
Prior to this we carry
out 10 iterations of spatial APE smearing
on the parallel transporters appearing
within eq.~(\ref{eq:gauss}) above~\cite{APE},
\footnotesize
\begin{equation}
U_{x,i}'=\mathrm{Proj}_{\mathop{\rm SU}(N)}
\left[\alpha\, U_{x,i}+\sum_{j\neq i}
\left(U_{x,j}U_{x+a\hat{\jmath},i}U_{x+a\hat{\imath},j}^\dagger
+U_{x-a\hat{\jmath},j}^\dagger
U_{x-a\hat{\jmath},i}
U_{x+a\hat{\imath}-a\hat{\jmath},j}\right)\right]\,,
\end{equation}
\normalsize
with smearing parameter $\alpha=2.5$, where
$U_\mu(x)$ is projected back into $\mathop{\rm SU}(N)$ after each iteration.

We invert the Dirac operator using the Conjugate Gradient
(CG) algorithm with even/odd preconditioning. We find
the number of CG iterations to be approximately independent of
$N$ at a given pion mass. Since each CG matrix-vector multiplication
has a cost proportional
to $N$ and we have to solve for $N$ different colour sources,
the total cost of producing a propagator is proportional to $N^2$.
Updating a gauge configuration (and also carrying out APE smearing)
involves matrix-matrix multiplications and so has a cost
proportional to $N^3$; however, for the relatively small values
of $N$ we use, the cost of calculating the propagators dominates,
so the total computer time spent on generating and analyzing
a fixed number of configurations is still
roughly proportional to $N^2$.

We find that the correlators become less noisy as $N$ increases,
so we are able to extract masses with smaller errors from the same
number of configurations. We find the errors for a given number
of configurations to decrease approximately like $1/\sqrt N$
(see tables~\ref{SU(2) masses}--\ref{SU(6) masses}).
Hence the number of configurations that would be required for
equal precision decreases like $1/N$ and the total cost is only
proportional to $N$. A heuristic argument supporting this
observed reduction in noise is the increase in the number of
degrees of freedom of the statistical system $\propto N^2$
at fixed volume. 

However, in~\cite{Lucini:2004my} it was observed that for glueballs
both the signal and the error decrease as $1/N^2$,
so the signal-to-noise ratio remains constant.
Hence arguments based only on the degrees of freedom can
be misleading and one must be more careful.
Repeating the argument of ref.~\cite{Lucini:2004my},
one finds that a crucial difference is that when
calculating the four-point correlator contributing to
the noise the intermediate states are dominated by two-meson states,
whereas in the case of glueballs the vacuum is a possible intermediate state.
The upshot of this is that for mesons the error does decrease
faster than the signal, by a factor of $1/N$,
in agreement with the heuristic argument above.
As noted above, the errors decrease by factors
$\simeq 1/\sqrt{N}$ that are somewhat smaller than this
best-case $1/N$ scenario.

\section{Results}
\label{sec:results}
\subsection{Extracting masses from correlation functions}
As described above, for each channel of interest we calculate
cross-correlation matrices
$C_{ij}(t)=\langle O_i(t) O_j^\dag(0) \rangle$, where $i,j$ correspond
to the size (`point', `narrow' or `wide') of the sources and sinks.
To extract masses, we first calculate eigenvectors $\psi_{t_0}^\alpha$
of the generalised eigenvalue problem~\cite{Michael:1985ne,Luscher:1990ck}:
\begin{equation}
C^{-1}(t_0)C(t_0+a) \psi_{t_0}^\alpha = \lambda_{t_0}^\alpha \psi_{t_0}^\alpha.
\end{equation}
We then project $C(t)$ onto the eigenvector $\psi^{\alpha}$ that corresponds to
the largest eigenvalue $\lambda^{\alpha}$
and perform periodic $\cosh$ fits, varying
the fit ranges, to extract the lowest mass.
We also vary $t_0$ within the range 0 -- $2a$; our results are
not very sensitive with respect to the value of $t_0$.
We estimate errors on
the masses using the jackknife method.

\EPSFIGURE{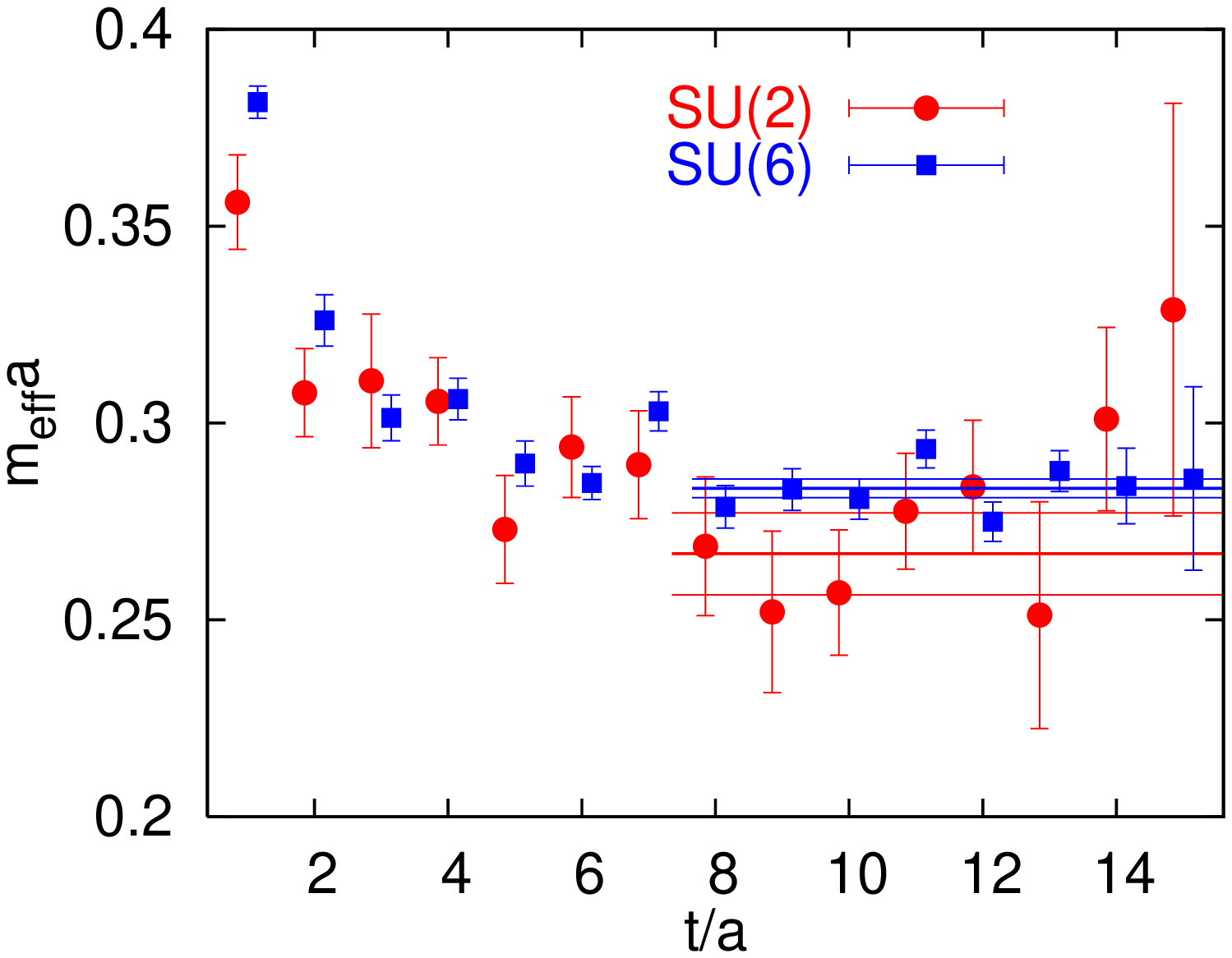,width=.85\textwidth}{
Effective masses and fits for the $\pi$ ground states at the
lightest mass-value for
$N=2$ ($\kappa=0.1510$, $m_{\pi}=1.29(5)\sqrt{\sigma}$)
and $N=6$ ($\kappa=0.15715$, $m_{\pi}=1.35(1)\sqrt{\sigma}$).
The lattice volumes were $16^3\times 32$ in both cases and
$t_0=a$ for $N=2$ while $t_0=2a$ for the more precise $\mathop{\rm SU}(6)$
data.
\label{effmass_fig}}

As an example we show in figure~\ref{effmass_fig} the
effective masses for our lightest pion in
$\mathop{\rm SU}(2)$ and $\mathop{\rm SU}(6)$. We see that there are clear plateaus
over a large range of $t$-values. These plots are typical
of those for other $N$ and other masses.

\EPSFIGURE{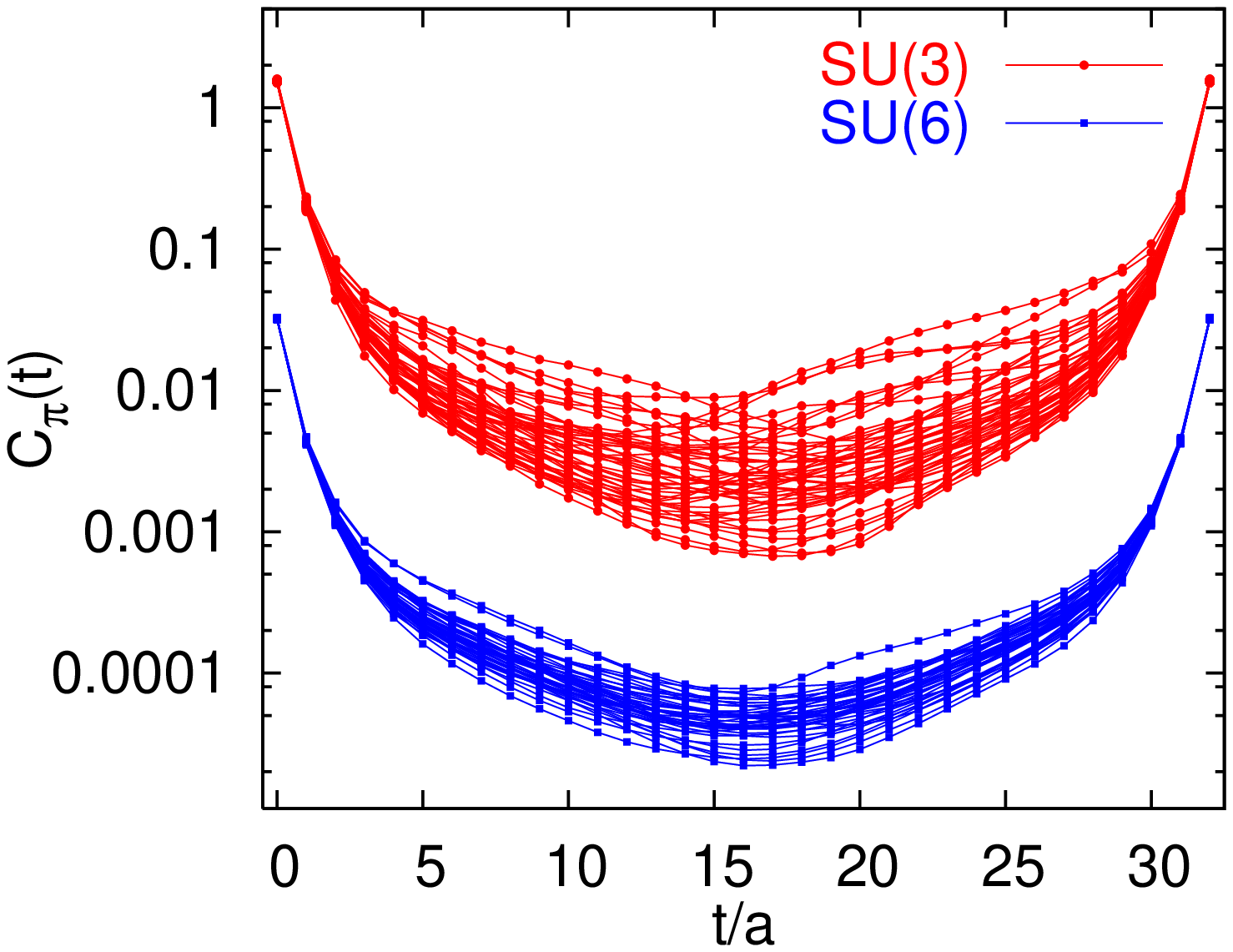,width=.85\textwidth}{
Point-point pseudoscalar correlators on individual $16^3\times 32$ gauge
configurations at the lightest mass values
in $\mathop{\rm SU}(3)$ ($\kappa=0.1547$, $m_{\pi}=1.33(2)\sqrt{\sigma}$)
and in $\mathop{\rm SU}(6)$ ($\kappa=0.15715$, $m_{\pi}=1.35(1)\sqrt{\sigma}$).
The $\mathop{\rm SU}(6)$ results have been shifted vertically for the comparison.
\label{noise_fig}}

As mentioned above (and as is also visible from figure~\ref{effmass_fig}),
we find that the correlators become less noisy, and hence the errors
on the masses decrease, as $N$ increases. We illustrate this in
figure~\ref{noise_fig}, where we compare point-point pseudoscalar
correlators on individual gauge configurations in
$\mathop{\rm SU}(3)$ and $\mathop{\rm SU}(6)$.
The pion masses are almost identical, $am_\pi=0.276(4)$ and $0.283(2)$,
respectively, but the scatter between individual configurations,
a measure of the noise, is about twice as large in $SU(3)$.
This is in agreement with the naive degrees of freedom argument
presented above.

\TABULAR{|c|c|c|c|c|c|}{\hline
$\kappa$ & Volume & $am_\pi$ & $m_\pi/\sqrt{\sigma}$ & $am_\rho$ & $m_\rho/\sqrt{\sigma}$ \\ \hline
0.1457 & $16^3\times32$ & 0.559( 4) & 2.67( 2) & 0.608( 8) & 2.90( 4) \\
0.1480 & $16^3\times32$ & 0.446( 5) & 2.13( 2) & 0.506(10) & 2.42( 5) \\
0.1500 & $16^3\times32$ & 0.336( 7) & 1.61( 3) & 0.428(14) & 2.04( 7) \\
0.1510 & $12^3\times32$ & 0.305(23) & 1.46(11) & 0.419(19) & 2.00( 9) \\
& $16^3\times32$ & 0.267(10) & 1.29( 5) & 0.392(20) & 1.87(10) \\
& $24^3\times32$ & 0.280( 3) & 1.34( 1) & 0.405( 8) & 1.93( 4) \\ 
& $\infty^3\times32$ & $0.279^{+2}_{-3}$ & 1.33( 1) & & \\ \hline
}{$\pi$ and $\rho$ masses in $\mathop{\rm SU}(2)$, in lattice units
and in units of the infinite-volume string tension. The infinite volume results
that are displayed for the pion in the last row are extrapolated.
\label{SU(2) masses}}

\TABULAR{|c|c|c|c|c|c|}{\hline
$\kappa$ & Volume & $am_\pi$ & $m_\pi/\sqrt{\sigma}$ & $am_\rho$ & $m_\rho/\sqrt{\sigma}$ \\ \hline
0.1500 & $16^3\times32$ & 0.552(3) & 2.64(2) & 0.606( 6) & 2.89(3) \\
0.1520 & $16^3\times32$ & 0.450(3) & 2.15(1) & 0.524( 5) & 2.50(2) \\
0.1537 & $16^3\times32$ & 0.347(2) & 1.66(1) & 0.448( 8) & 2.14(4) \\
0.1547 & $12^3\times32$ & 0.306(9) & 1.46(4) & 0.432(14) & 2.06(7) \\
& $16^3\times32$ & 0.276(4) & 1.32(2) & 0.398(10) & 1.90(5) \\
& $\infty^3\times32$ & $0.274^{+4}_{-9}$ & $1.31^{+2}_{-4}$ & & \\ \hline
}
{
The same as table~\protect\ref{SU(2) masses} for
$\mathop{\rm SU}(3)$.
\label{SU(3) masses}}

\TABULAR{|c|c|c|c|c|c|}{\hline
$\kappa$ & Volume & $am_\pi$ & $m_\pi/\sqrt{\sigma}$ & $am_\rho$ & $m_\rho/\sqrt{\sigma}$ \\ \hline
0.1520 & $16^3\times32$ & 0.544(2) & 2.60(1) & 0.600( 3) & 2.87(2) \\
0.1540 & $16^3\times32$ & 0.438(2) & 2.09(1) & 0.519( 4) & 2.48(2) \\
0.1554 & $16^3\times32$ & 0.354(3) & 1.69(1) & 0.459( 5) & 2.19(2) \\
0.15625 & $12^3\times32$ & 0.284(7) & 1.36(3) & 0.402(13) & 1.92(6) \\
& $16^3\times32$ & 0.295(3) & 1.41(1) & 0.424( 7) & 2.02(3) \\
& $\infty^3\times32$ & $0.294^{+3}_{-7}$ & $1.40^{+1}_{-3}$ & & \\ \hline
}
{
The same as table~\protect\ref{SU(2) masses} for
$\mathop{\rm SU}(4)$.
\label{SU(4) masses}}

\TABULAR{|c|c|c|c|c|c|}{\hline
$\kappa$ & Volume & $am_\pi$ & $m_\pi/\sqrt{\sigma}$ & $am_\rho$ & $m_\rho/\sqrt{\sigma}$ \\ \hline
0.1525 & $16^3\times32$ & 0.558(2) & 2.67(1) & 0.618(2) & 2.95(1) \\
0.1550 & $16^3\times32$ & 0.426(2) & 2.04(1) & 0.507(4) & 2.42(2) \\
0.1563 & $16^3\times32$ & 0.345(2) & 1.65(1) & 0.447(5) & 2.14(2) \\
0.15715 & $12^3\times32$ & 0.297(4) & 1.42(2) & 0.420(9) & 2.01(4) \\
& $16^3\times32$ & 0.283(2) & 1.35(1) & 0.416(6) & 1.99(3) \\
& $\infty^3\times32$ & $0.282^{+3}_{-8}$ & $1.35^{+1}_{-4}$ & & \\ \hline
}
{
The same as table~\protect\ref{SU(2) masses} for
$\mathop{\rm SU}(6)$.
\label{SU(6) masses}}

We have calculated correlators for every quark bilinear $J^{PC}$ channel.
However, we find that only the $J^{PC}=0^{-+}$ channel, corresponding to
the pion, and the $J^{PC}=1^{--}$ channel, corresponding to the $\rho$ meson,
have sufficiently strong signals to accurately extract masses at
present statistics. We present
the masses of the ground states in these channels in
tables~\ref{SU(2) masses}--\ref{SU(6) masses}.

\subsection{Finite size effects}
Finite volume corrections are expected to be largest at the smallest
pion masses, and at the smallest values of $N$. Comparing our
results for the lightest $\pi$ masses between different volumes
(tables~\ref{SU(2) masses}--\ref{SU(6) masses}), we see that the
differences indeed appear to decrease with $N$. We also see that the
corrections for the $\rho$ meson are smaller than for the $\pi$,
as expected since the $\rho$ is heavier.

We can estimate the infinite-volume masses by fitting our finite
volume results to the form,
\begin{equation}
m_\pi(L) = m_\pi(\infty) \left[ 1 + C\,\exp(- m_\pi(\infty) L)\right],
\label{finite size correction}
\end{equation}
with $C$ positive~\cite{Colangelo1,Colangelo2}.
For $\mathop{\rm SU}(2)$ we have three values of $L$ with which
to carry out the fit, and we obtain $am_\pi(\infty)=0.279^{+2}_{-3}$
and $C=0.6^{+2.1}_{-0.6}$.
Note that the data can also be fitted to a constant.
For the other values of $N$ we only have two lattice sizes
for a two-parameter fit, and furthermore the $L=12$ lattices
are rather small so higher order corrections to
eq.~(\ref{finite size correction}) could be substantial.
Hence for these cases we take the conservative approach
\footnote{Note that $C\rightarrow 0$ as $N\rightarrow\infty$.}
of assuming that the coefficient $C$ remains constant at
its $\mathop{\rm SU}(2)$ value for all $\mathop{\rm SU}(N)$.
We have included the infinite-volume pion masses obtained from
these fits in tables~\ref{SU(2) masses}--\ref{SU(6) masses}.
In no case do we obtain any statistically significant deviation
between the infinite volume extrapolated result and those
obtained on our $L=16a\approx 1.5$~fm lattices.

In principle we should also carry out similar corrections
to the $\rho$ meson masses. However, the corrections are
much smaller for the $\rho$, and will be negligible compared to our
statistical errors. So for the $\rho$ we simply use the results from
our largest volumes. Similarly, at larger masses it is not necessary
to include finite-volume effects since they will be smaller than
our statistical errors, for both the $\pi$ and the $\rho$.

We note that our chiral extrapolations, described below,
are not very sensitive to the details of our finite-volume corrections.
This is because the chiral fits are mostly controlled by the data at
higher mass
values, which have significantly smaller errors, and are much less
sensitive to potential finite-volume corrections.

\subsection{Determination of the critical hopping parameter}
\subsubsection{$\kappa_c$ at finite $N$}
\EPSFIGURE{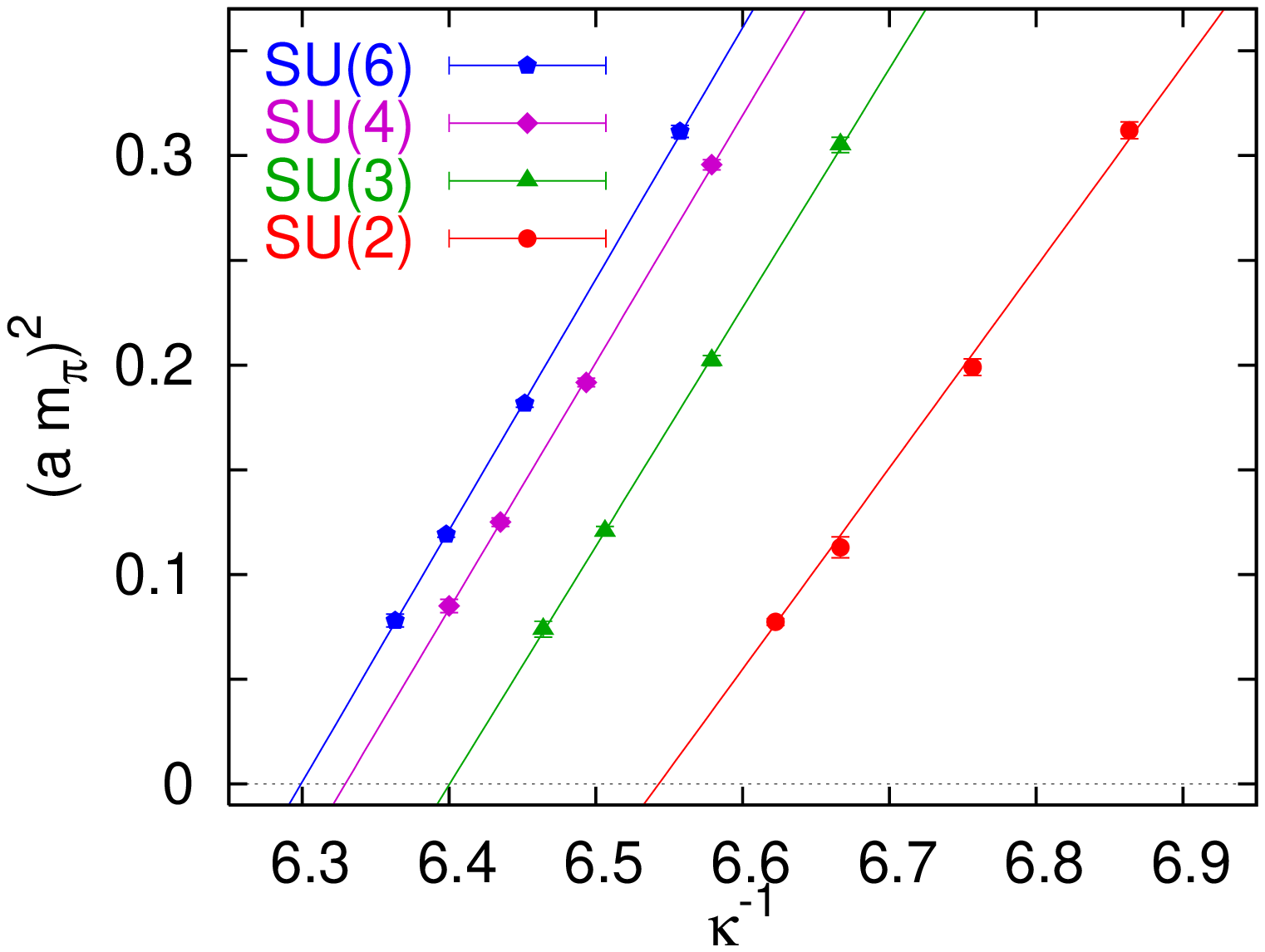,width=.85\textwidth}
{$(a m_\pi)^2$ as a function of $1/\kappa$,
eq.~(\protect\ref{pion mass equation}).
\label{pionfit_fig}}

\TABULAR{|c|l|c|}{\hline
$N$ & $\kappa_c$ & $\chi^2/n$ \\ \hline
2 & 0.1528(1)(13)& 3.0 \\
3 & 0.1562(1)( 3) & 0.9 \\
4 & 0.1580(1)( 4) & 1.7 \\
6 & 0.1588(1)( 2) & 1.7 \\
$\infty$&0.1596(2)&0.3\\\hline
}{
The critical hopping parameter $\kappa_c$, with the
reduced $\chi^2$ of the fits. The first errors are statistical,
the second systematic.
\label{kappa_table}}

The Wilson action quark mass undergoes an additive
renormalization $(2a\kappa_c)^{-1}$, see eq.~(\ref{bare mass}).
Thus we expect that the pion mass will be related to $\kappa$ 
(up to quenched chiral logs) by,
\begin{equation}
(am_\pi)^2=A \left(\frac{1}{\kappa}-\frac{1}{\kappa_c}\right)\,.
\label{pion mass equation}
\end{equation}
By fitting to this equation we can extract $\kappa_c$ for each $N$.
We obtain good fits for each $N$, which we display in figure~\ref{pionfit_fig}.
Note also that the pion masses are horizontally aligned across
the different $N$-values, indicating that we have succeeded
to approximately match the $\kappa$-values to lines
of constant physics,
thus eliminating another possible source of systematic
bias. The values of $\kappa_c$ that
we obtain are shown in table~\ref{kappa_table}.
We note that while the lattice 't Hooft coupling $2N^2/\beta$ at fixed
string tension decreases
with increasing $N$ (see table~\ref{parameters}), the $\kappa_c$-values
move away from the free field limit $\kappa_{c,{\rm free}}=0.125$.

We are dealing with the quenched theory
and hence eq.~(\ref{pion mass equation}) should be modified
by quenched chiral logarithms~\cite{Sharpe}, giving 
\begin{equation}
(am_\pi)^2=A' \left( {\frac{1}{\kappa}-\frac{1}{\kappa_c}}\right)^{1+\delta}\,.
\end{equation}

We also fitted our results to this parametrization but have
found $\delta$ to be consistent with zero, confirming that at
$m_{\pi}>1.3\sqrt{\sigma}\approx 580$~MeV we are not yet sensitive
to quenched chiral logarithms. 
Comparing our best fits using eq.~(\ref{pion mass equation}) to those
where we allow $\delta$ to vary, we estimate the uncertainty
on $\kappa_c$ due to the lack of control over the chiral logs
to be significantly larger than
our statistical errors (see table~\ref{kappa_table}).
This error should decrease with $N$ since
the quenched theory will become equivalent to the unquenched theory
in the large-$N$ limit. Indeed, the $N=2$ data allow for more
curvature than the $N>2$ data sets.

\subsubsection{$\kappa_c$ in the large-$N$ limit}
\EPSFIGURE[ht]{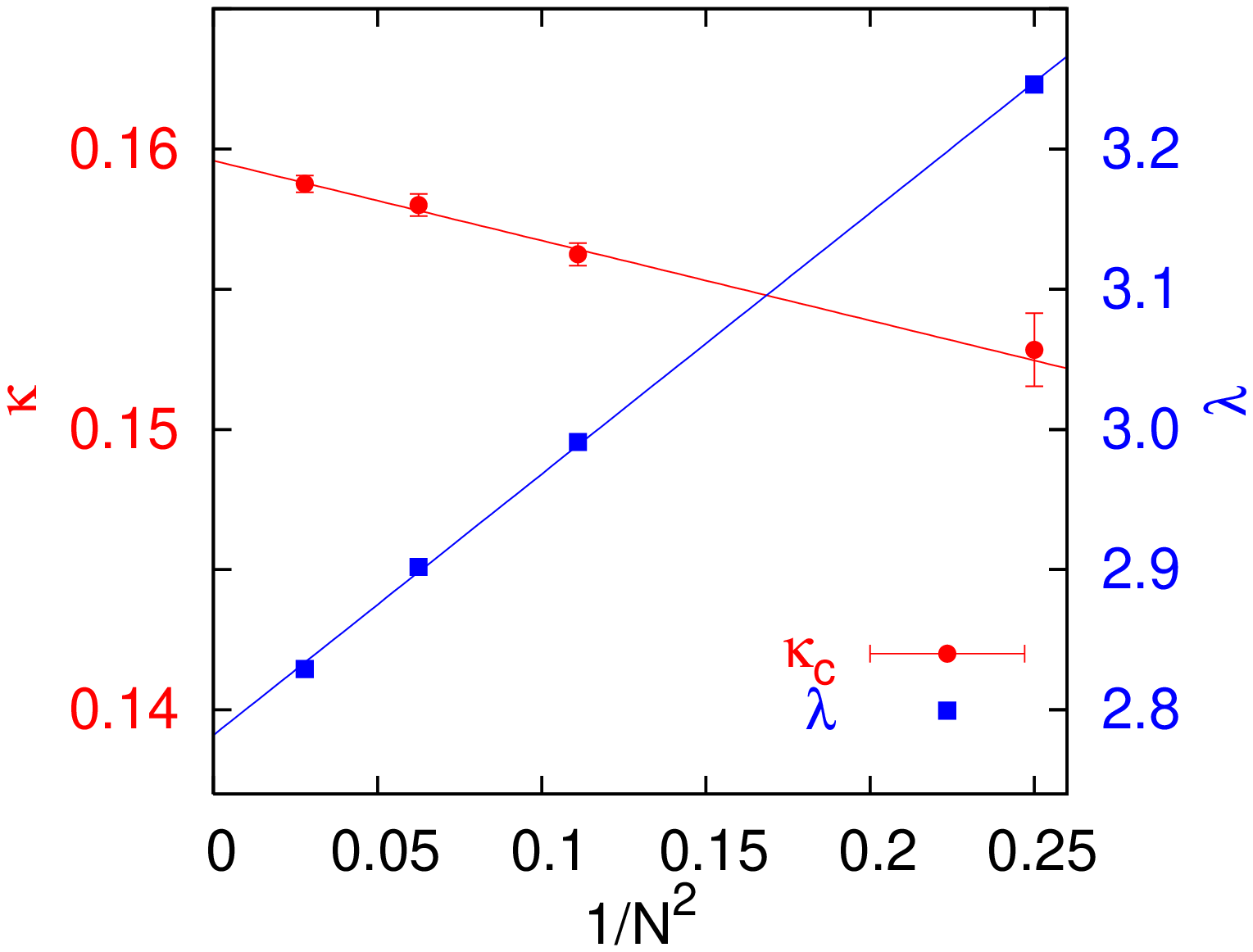,width=.85\textwidth}
{$\kappa_c$ as a function of $1/N^2$. For comparison we
also show the infinite $N$ extrapolation of the
lattice 't Hooft coupling $2N^2/\beta$.
\label{kappafit_fig}}

We expect $\kappa_c$ to have $\mathcal{O}(1/{N^2})$
correction to its large-$N$ value. Hence we fit the values
of table~\ref{kappa_table} to the form,
\begin{equation}
\kappa_c=\kappa_c(N=\infty)+\frac{c}{N^2}\,.
\label{kappa_c equation}
\end{equation}
After including the systematic uncertainties from the chiral extrapolation
we obtain values $\kappa_c(\infty)=0.1596(2)$ and $c=-0.028(3)$.
Some of the systematics will be correlated and we obtain a rather
small $\chi^2/n\approx 0.27$. We display the $1/N^2$ extrapolation
of $\kappa_c$ in
figure~\ref{kappafit_fig}. We also include the $1/N^2$
extrapolation of the lattice 't Hooft coupling at fixed string tension.
Note that this extrapolates to the value $\lambda(N=\infty)=2.780(4)$.

\subsection{The $\rho$ meson mass}
\subsubsection{$m_{\rho}$ at finite $N$}
\EPSFIGURE{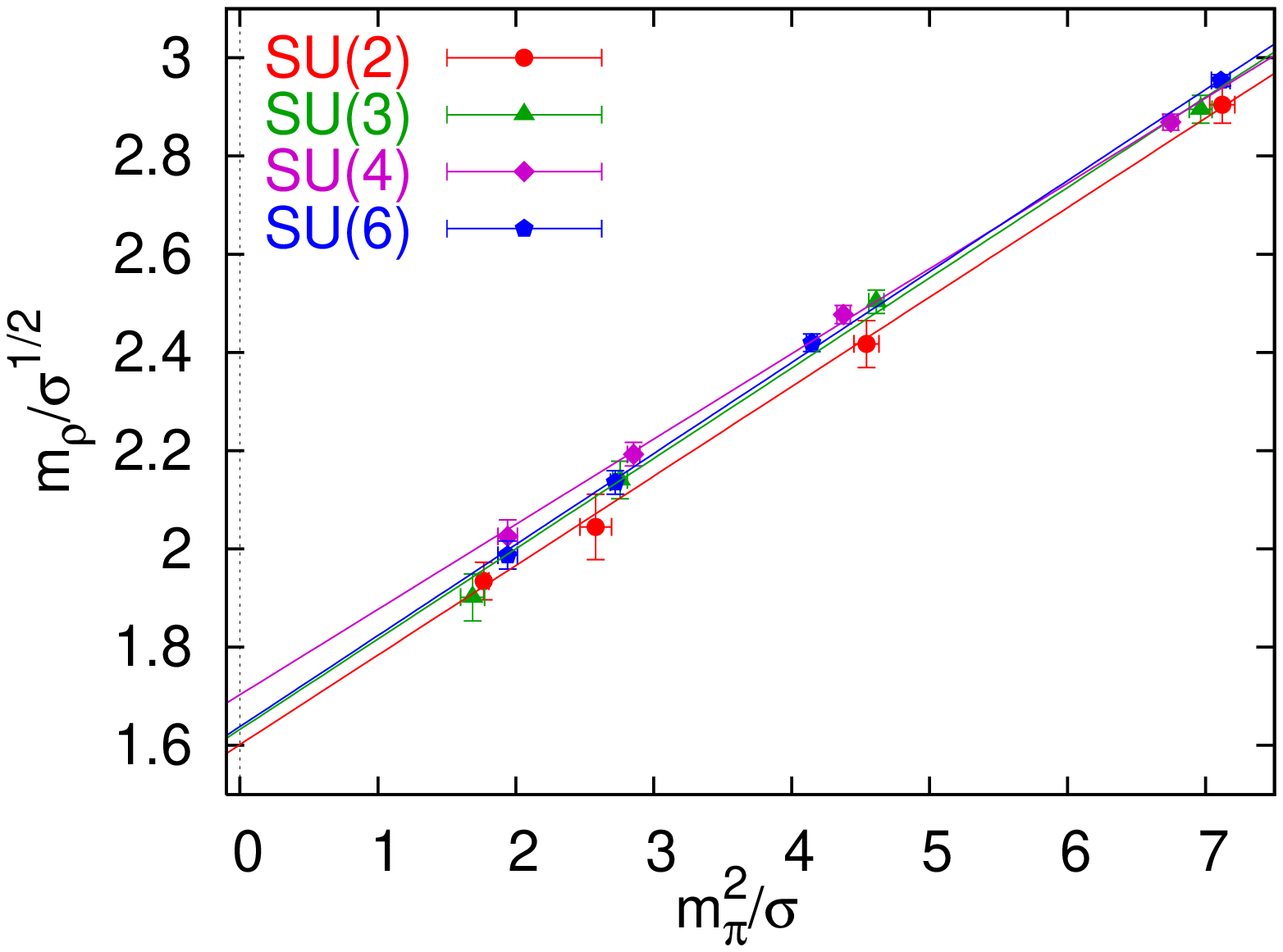,width=.85\textwidth}
{Fits of $m_\rho/\sqrt{\sigma}$ as functions of $m_\pi^2/\sigma$,
eq.~(\protect\ref{rho mass equation}), at different $N$.
\label{rhofit_fig}}

\TABULAR{|c|c|c|c|}{\hline
$N$ & $m_\rho(0)/\sqrt{\sigma}$ & $B$ & $\chi^2/n$ \\ \hline
2 & 1.60(5) & 0.182(10) & 0.2 \\
3 & 1.63(4) & 0.184( 9) & 1.0 \\
4 & 1.70(3) & 0.174( 6) & 0.5 \\
6 & 1.64(3) & 0.185( 4) & 0.6 \\ \hline
}{
The fit parameters of
eq.~(\protect\ref{rho mass equation}) for each $N$, with the reduced
$\chi^2$-values.
\label{rho_table}}

In chiral perturbation theory, as well as in the heavy quark
limit, $m_\rho$ depends linearly on the quark mass $m_q$.
Within our range of pion masses $m_{\pi}/\sqrt{\sigma}
\approx 1.3\ldots 2.6$ we find
$m_\pi^2$ to linearly depend on $\kappa^{-1}$ and
therefore to be proportional to the quark mass.
Hence, we can fit our $\rho$ masses to the
parametrization,
\begin{equation}
\frac{m_\rho}{\sqrt{\sigma}}=\frac{m_\rho(0)}{\sqrt{\sigma}} + B \frac{m_\pi^2}{\sigma}\,,
\label{rho mass equation}
\end{equation}
where $m_\rho(0)$ denotes the $\rho$ meson mass in the chiral limit.
Quenched chiral logarithms will modify the relationship between $m_q$ and $m_\pi^2$,
however~eq.~(\ref{rho mass equation}) remains valid~\cite{Sharpe}.

We obtain good fits in each case,
which we show in figure~\ref{rhofit_fig}. We
display corresponding parameter values $m_\rho(0)/\sqrt{\sigma}$ and $B$
in table~\ref{rho_table}.

\subsubsection{$m_{\rho}$ in the large-$N$ limit}
\EPSFIGURE{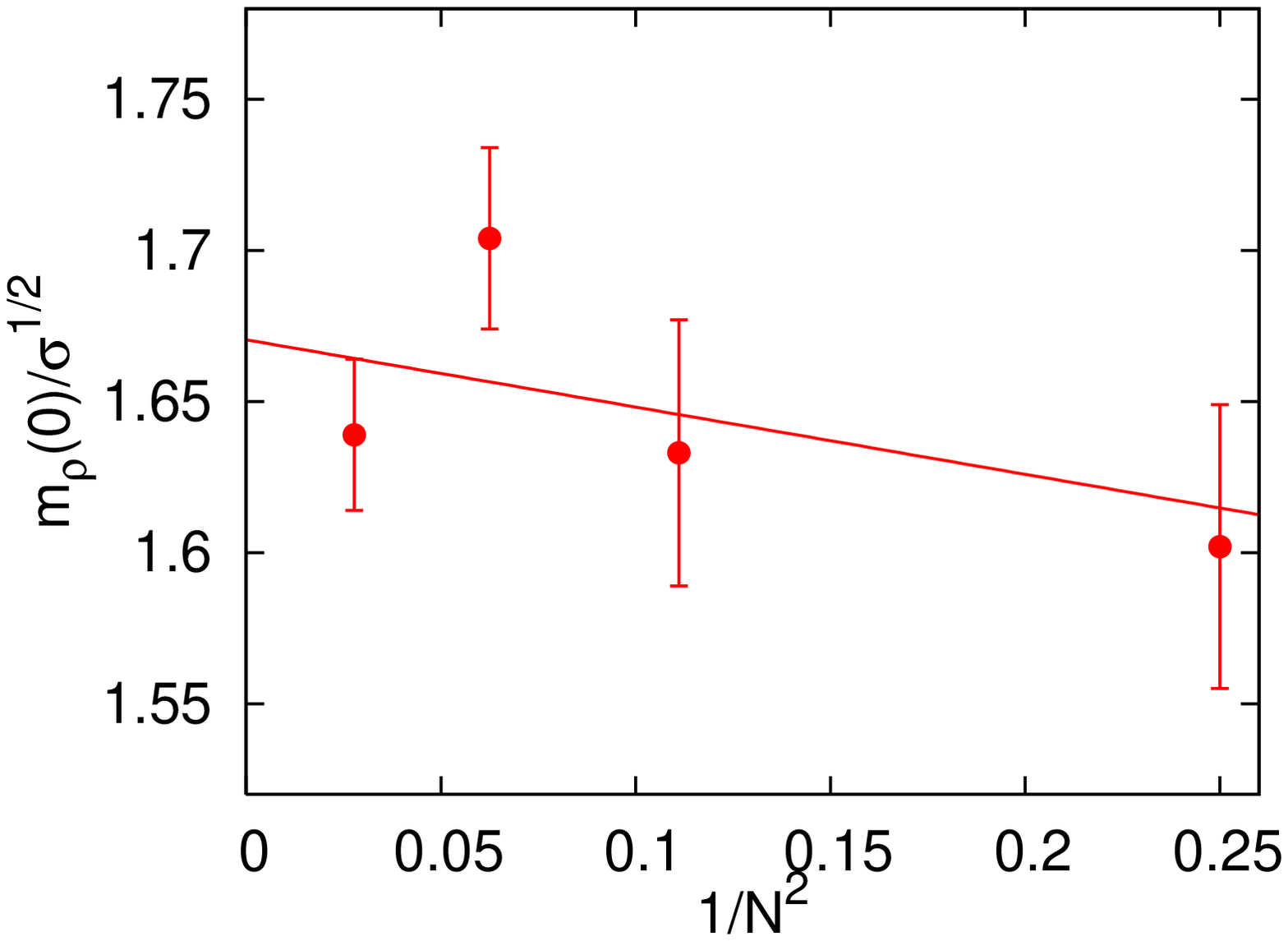,width=.85\textwidth}
{$m_\rho/\sqrt{\sigma}$ in the chiral limit,
as a function of $1/N^2$.
\label{m0fit_fig}}

In the quenched approximation
the parameters $m_\rho(0)/\sqrt{\sigma}$ and $B$ are each expected
to have $\mathcal{O}(1/{N^2})$ corrections to their large-$N$ values,
in analogy to eq.~(\ref{kappa_c equation}). We see from
figure~\ref{rhofit_fig} that the results and fits for different $N$
are all very similar, with all points nearly lying on one line,
so these corrections should be small.
We obtain,
\begin{equation}
m_\rho(0)/\sqrt{\sigma}=1.670(24)-\frac{0.22(23)}{N^2}\,,
\label{mrho0 equation}
\end{equation}
and
\begin{equation}
B=0.182(5)-\frac{0.01(5)}{N^2}\,.
\label{B equation}
\end{equation}
The fits are both good, with reduced $\chi^2$-values of 1.8 and 1.3,
respectively. We plot the fit for
$m_\rho(0)/\sqrt{\sigma}$ in figure~\ref{m0fit_fig}.
Note that the $N$-dependence is extremely small both for
$m_{\rho}/\sqrt{\sigma}$ and for $B$.
In particular, the difference between
$m_\rho(0)$ for $\mathop{\rm SU}(3)$ and $\mathop{\rm SU}(\infty)$
amounts to only $(1.4\pm1.6)$~\%.

Equating $\sqrt{\sigma}\approx 444$~MeV, 
the large-$N$ chiral $\rho$ mass, eq.~(\ref{mrho0 equation}), corresponds to,
\begin{equation}
m_\rho(m_{\pi}=0,N=\infty) \approx 741(11)\, \mathrm{MeV}\,.
\end{equation}
The above value $\sigma\approx 1\,\mbox{GeV}/\mbox{fm}$ is motivated
by Regge trajectories, potential model fits to quarkonium spectra and
unquenched $N=3$ lattice data~\cite{Bali:2000vr}.
However, there is a systematic uncertainty
associated with it, in particular also since the QCD string
at finite $N$ and $n_F>0$ will decay, once a
critical distance is reached~\cite{Bali:2005fu}. Moreover,
no $N=\infty$ experiment exists.
Nonetheless, we find it interesting
that this mass-prediction for the
stable $N=\infty$ $\rho$ meson comes close to the mass
$m_{\rho}\approx 775$~MeV of
the experimental $\rho$ resonance, which in fact has a significant
decay width, $\Gamma\approx 150$~MeV.
Note that at physical $\pi$ mass we obtain
the large-$N$ limit $m_{\rho}\approx 749(11)$~MeV.
The (unknown) scale uncertainty from the string tension
is not included in the above error
estimates and neither are finite lattice spacing effects.
The continuum limit extrapolation of
section~\ref{sec:compare} below increases
our large-$N$ $m_{\rho}$ prediction
by another $\approx 45$~MeV.

\subsection{Excited states}
In principle, our correlation matrices enable us to extract the
masses of excited states as well as of the ground states. However,
in practice we obtain poor mass plateaus for the excited states,
and our statistical errors are rather large. Having said that, our
results are consistent with the first excited pseudoscalars and
vectors having no $N$-dependence at the 10~\% level. Also, we find 
that our correlators at a given pion mass are similar for all $N$,
not just at large $t$ where they are dominated by the ground state,
but all the way back to $t=0$ where many excited states contribute.
An example of this can be seen in figure~\ref{noise_fig}. This is
consistent with the entire spectrum of excited states to only weakly
depend on $N$. The correlators also depend on the overlaps of our
operators with the excited states, and these overlaps in turn
depend on wavefunctions of the excited states.
So this also suggests that the meson sizes
and internal properties do not change strongly
with $N$.

\section{The continuum limit}
\label{sec:compare}
Presumably there are
lattice spacing corrections to our results. We can get
some idea of their size from the study
of Del Debbio \emph{et al.}~\cite{DelDebbio:2007wk},
on somewhat coarser lattices.

Turning first to the large-$N$ limit of their results, and
using the facts that their lattice spacing is given by
$a\approx 1/(5\,T_c)$, where $T_c$ denotes the de-confinement temperature,
and that at $N=\infty$ one
has $T_c/\sqrt{\sigma}\approx 0.5970$~\cite{Lucini:2005vg}, they obtain,
\begin{equation}
\frac{m_\rho}{\sqrt{\sigma}}=1.609(9) + 0.1750(3) \frac{m_\pi^2}{\sigma}\quad; \qquad a\sqrt{\sigma}=0.3350\,.
\end{equation}
We can compare this to our large-$N$ result,
eqs.~(\ref{rho mass equation})--(\ref{B equation}),
\begin{equation}
\frac{m_\rho}{\sqrt{\sigma}}=1.670(24) + 0.182(5) \frac{m_\pi^2}{\sigma}\quad; \qquad a\sqrt{\sigma}=0.2093\,.
\label{our large N limit}
\end{equation}
We see that, despite the 60~\% difference in lattice spacings,
the results for both the chiral limit and the slope of the
$\rho$ mass are very similar, differing by less than 5~\%.
Also the ratio of our and their results remains remarkably independent of the
$\pi$ mass, when expressed in units of $\sigma$.

The leading lattice artefacts are of
order $a$ and a simple linear extrapolation yields,
\begin{equation}
\frac{m_\rho}{\sqrt{\sigma}}=1.77(5) + 0.193(14) \frac{m_\pi^2}{\sigma}\quad; \qquad (a\rightarrow 0)\,.
\label{eq:contin}
\end{equation}
Obviously, with just two values of the lattice spacing,
we have little control over the systematics of this
continuum limit extrapolation. Therefore we also attempt
a purely quadratic extrapolation in $a$ and find the central values
1.70 and 0.186 for the two above parameters, respectively.
Based on this and results of previous large scale
$N=3$ quenched spectroscopy studies with Wilson action~\cite{Aoki:1999yr},
that included similar lattice spacings (as well as finer ones),
we conclude that the statistical errors stated
above are sufficiently large to accommodate the possible
contributions from
subleading terms.

\EPSFIGURE{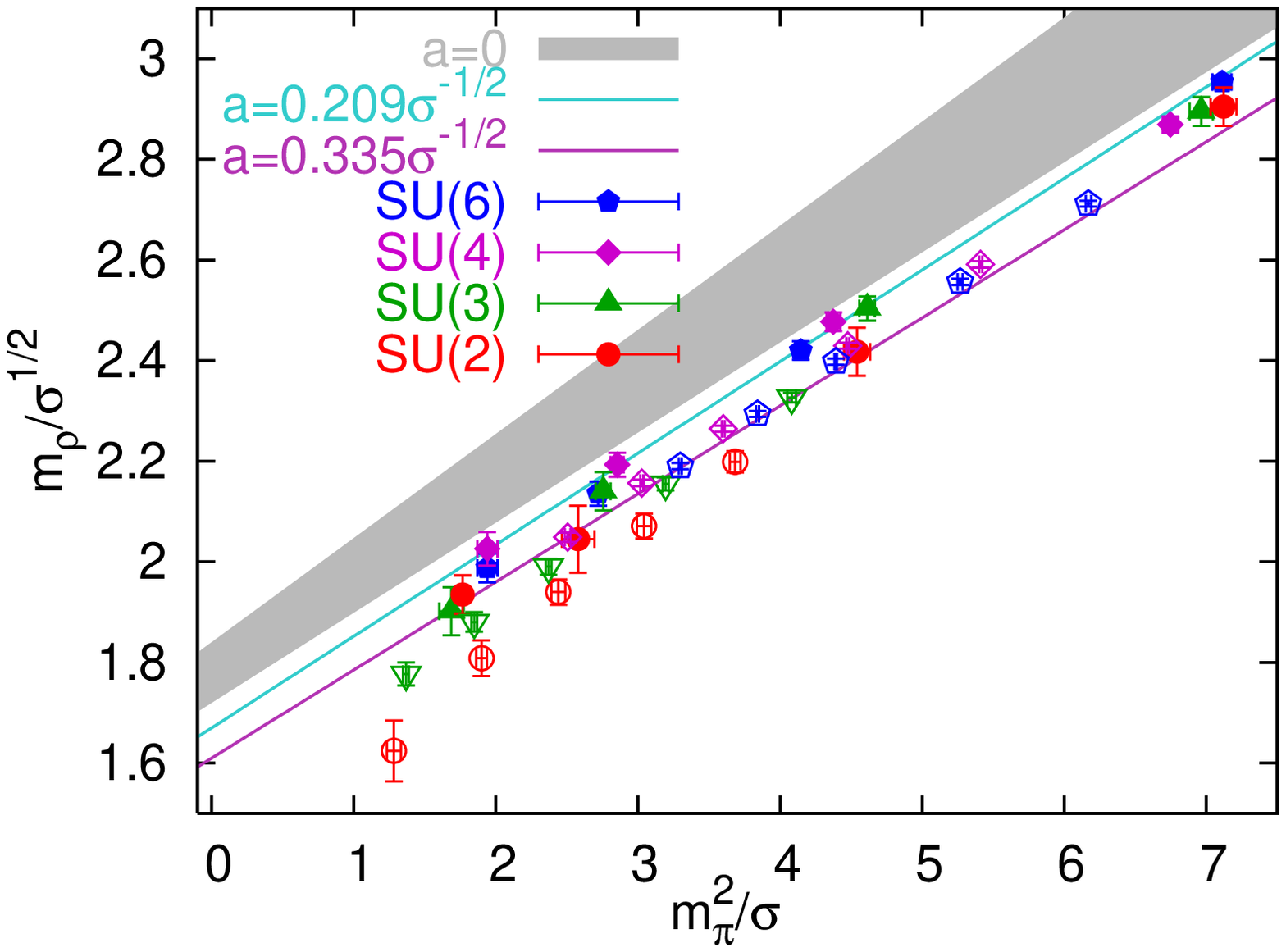,width=.96\textwidth}{Results
for $m_\rho/\sqrt{\sigma}$ and $m_\pi^2/\sigma$ from this work (full symbols,
$a\approx 0.209/\sqrt{\sigma}$)
and from ref.~\protect\cite{DelDebbio:2007wk} (open symbols,
$a=0.2/T_c\approx 0.335/\sqrt{\sigma}$).
The lines correspond to the respective $N\rightarrow\infty$ limits
and the grey error band denotes our large-$N$ continuum limit
estimate eq.~(\protect\ref{eq:contin}).
\label{comparison_fig}}

We now turn to the finite-$N$ corrections. Here the
two sets of results appear to be rather different. Our results show a
very weak $N$-dependence, with for example the difference between
$m_\rho(0)$ for $\mathop{\rm SU}(3)$
and $\mathop{\rm SU}(\infty)$ being only $(1.4\pm 1.6)$~\%.
The corresponding value of ref.~\cite{DelDebbio:2007wk} is $(12.8\pm 0.6)$~\%,
much larger. Most of this difference can be attributed to the different
ways of setting the scale;
the ratio $T_c/\sqrt{\sigma}$ depends on $N$~\cite{Lucini:2005vg}. To
remove this effect we have rescaled the results of
ref.~\cite{DelDebbio:2007wk},
using the appropriate values of $T_c/\sqrt{\sigma}$ for each
$N$~\cite{Lucini:2005vg}. We compile all
these results in figure~\ref{comparison_fig}.
Again we remark that the results of ref.~\cite{DelDebbio:2007wk}
were obtained at different values of $a\sqrt{\sigma}$ for the
different $N$. Also note that the range of pion masses
covered in this case varies with $N$.
Nevertheless, we see that when plotted in this way the
two studies agree very closely,
with the exception of $\mathop{\rm SU}(2)$.
This suggests that most of the
difference between the two data sets is due to the different
methods of setting the scale, rather than from lattice corrections.
The finite lattice spacing effects might be larger for $\mathop{\rm SU}(2)$,
accounting for the deviation there. 

The fact that
the $N$-dependence at our finer lattices is already consistent
with zero leaves little room for $1/N^2$ corrections to
$m_{\rho}/\sqrt{\sigma}$ being larger than $3~\%$ at $N=3$, at least for
pion masses $m_{\pi}/\sqrt{\sigma}<2.6$.
The figure also contains the large-$N$ extrapolations of the
data sets at the two lattice spacings and our continuum limit
estimate eq.~(\ref{eq:contin}) (grey error band).

All this suggests that our finite-$a$ results deviate by
no more than 5--10~\%
from the continuum limit. This is true both for the
large-$N$ limit, eq.~(\ref{our large N limit}), and for the
finite-$N$ corrections.
Also, these corrections are smaller when the masses are expressed
in terms of the string tension than when they are expressed in terms
of the de-confinement temperature. Note that this is opposite
to the situation with respect to the scalar glueball
which scales better between different $N$ when normalized with
respect to $T_c$,
rather than by $\sqrt{\sigma}$~\cite{Lucini:2004my}.

\section{Summary and Discussion}
\label{sec:discuss}
We have studied the $\rho$ and $\pi$ meson masses
in quenched $\mathop{\rm SU}(N)$ QCD with $N=2,3,4,6$
at one value of the lattice spacing $a\sqrt{\sigma}= 0.2093$,
in units of the
string tension, at four different values of the quark
mass. Our main result can be
summarized by combining
eqs.~(\ref{rho mass equation}),~(\ref{mrho0 equation})
and~(\ref{B equation}), giving
\begin{equation}
\frac{m_\rho(m_{\pi})}{\sqrt{\sigma}}=1.670(24)-\frac{0.22(23)}{N^2} + \left(0.182(5)-\frac{0.01(5)}{N^2}\right)\frac{m_\pi^2}{\sigma}\,.
\end{equation}
Combining our data with that of
Del Debbio \emph{et al.}~\cite{DelDebbio:2007wk}, allows us to
extrapolate the large-$N$ result to the continuum limit,
see eq.~(\ref{eq:contin}).

We can compare our results to predictions
from AdS/QCD correspondence. As an example, we consider the
case of the Constable-Myers deformation~\cite{Constable}
analyzed in ref.~\cite{Erdmenger}.
In their original units of $m_{\rho}(0)$ this reads~\cite{Evans},
\begin{equation}
\frac{m_\rho(m_{\pi})}{m_{\rho}(0)}\approx 1+0.307\left(\frac{m_\pi}{m_{\rho}(0)}
\right)^2\,.
\label{eq:erd}
\end{equation}
At our value of $a$ we obtain 0.304(3) for this slope in units
of $m_{\rho}(0)$, while
the values obtained in ref.~\cite{DelDebbio:2007wk} can be translated
into a coefficient $0.2816(2)$. A linear continuum limit extrapolation
of these two numbers yields 0.341(4) and a quadratic extrapolation
results in 0.318(3). The expected dominant behaviour is linear
and hence we quote $0.341\pm 0.023$ as our large-$N$ continuum
limit result, where the error is dominated by the systematics of the
continuum limit extrapolation. The above AdS/QCD prediction eq.~(\ref{eq:erd})
agrees reasonably well with this result.

In the large-$N$ limit the quenched theory is equivalent to the full,
unquenched, theory. We have found that the $\mathcal{O}(1/N^2)$ quenched corrections to
the large-$N$ result are very small. What about the
$\mathcal{O}(n_F/N)$ corrections
of the unquenched theory? We can get an idea of
their size without needing to carry out full unquenched simulations,
by comparing our results with experiment. Converting
the
$a\rightarrow 0$ and $N\rightarrow\infty$ extrapolated result of
eq.~(\ref{eq:contin}) into physical units gives,
\begin{equation}
m_{\rho}(m_{\pi})=(786\pm 22)\,\mbox{MeV}+(435\pm 34)\frac{m_{\pi}^2}{\mbox{GeV}}\,.
\end{equation}
At physical $\pi$ mass this yields an infinite $N$ value,
$m_{\rho}=(794\pm 23)$~MeV, in perfect agreement with
experiment: $m_{\rho}\approx 775$~MeV.
This might suggest that not only the quenched
$1/N^2$-corrections but also the unquenched $n_F/N$-corrections
are small at $N=3$.
 
Of course the string tension value $\sqrt{\sigma}=444$~MeV
is arbitrary and we did not associate any error to this
choice. Still this is very encouraging and might
explain the success of the quenched approximation
in light hadron spectroscopy calculations
of the 1980s and 90s~\cite{Aoki:1999yr}.
It is certainly worthwhile to extend the present study to
other mesonic channels, to confirm this picture.

As discussed above, the fact that our results agree reasonably well
with those of ref.~\cite{DelDebbio:2007wk} although our lattice
spacings differ by as much as 60~\% suggests that the systematic
error due to the finite lattice spacing is already small at
$a\approx 0.093$~fm for the Wilson action. Our continuum limit
extrapolation confirms it to be smaller than 10~\%.
It would be nice to further constrain the continuum limit
by repeating our work with the same analysis methods
at another, smaller, lattice spacing.

The errors associated with
the neglect of quenched chiral logs in our pion mass fits
could be reduced by going to lighter $\pi$ masses; the obstacle of
exceptional configurations should be reduced at high $N$ since the
Dirac matrix eigenvalue distribution becomes
narrower~\cite{Bali:2007kt}. Our calculations have been carried
out on more than one volume, so we already have some control over
the small finite volume corrections. These should in any case
disappear in the large-$N$ limit.

We have not been able to obtain accurate masses for excited states
in this work, although our results do suggest that they do not have
a strong $N$-dependence. We intend to use improved meson
operators to calculate the masses of the low-lying excited states
more accurately in the future. This should yield several independent
mass ratios in the large-$N$ limit which can be used to constrain
AdS/QCD models. This would also shed more
light on the question if the use of the quenched approximation at
$n_F/N=3/3$
is justifiable from a large-$N$ viewpoint and relate low energy
constants in chiral Lagrangians.
We are also planning to study the scalar sector and flavour singlet
diagrams.

\acknowledgments{We thank Johanna Erdmenger, Biagio Lucini and Mike Teper
for discussions. The computations were mainly performed on the
Regensburg QCDOC machine and we thank Stefan Solbrig
for keeping this alive. This work is supported by the
EC Hadron Physics I3 Contract RII3-CT-2004-506087 and by the
GSI University Program Contract RSCHAE. F.B. is supported by the STFC.}

\end{document}